\documentstyle[iopconf1,epsf,psfig]{article}
\newcommand{\beq}{\begin{equation}}

\newcommand{\eeq}{\end{equation}}
\newcommand{\bea}{\begin{eqnarray}}
\newcommand{\eea}{\end{eqnarray}}
\newcommand{\F}{\Phi}
\newcommand{\f}{\phi}
\newcommand{\vf}{\varphi}
\newcommand{\Q}{\tilde{Q}_{_L}}
\newcommand{\q}{\tilde{q}_{_R}}
\newcommand{\Lp}{\tilde{L}_{_L}}
\newcommand{\lp}{\tilde{l}_{_R}}

\begin{document}

\title{
Superball dark matter\footnote{Invited talk at Second International
  Conference on Dark Matter in Astrophysics and Particle Physics (DARK-98),
  Heidelberg, Germany, July  20-25, 1998}
}

\author{Alexander Kusenko\footnote{E-mail:
Alexander.Kusenko@cern.ch}}

\affil{Theory Division, CERN, CH-1211 Geneva, Switzerland
}

\beginabstract

Supersymmetric models predict a natural dark-matter candidate, stable 
baryonic Q-balls.  They could be copiously produced in the early Universe
as a by-product of  the Affleck-Dine baryogenesis.  
I review the cosmological and astrophysical implications, methods of
detection, and the present limits on this form of dark matter. 

\endabstract

\section{Introduction}

Non-topological solitons associated with some conserved global
charge (Q-balls) appear in scalar field theories that have some
``attractive'' interactions~\cite{nts,q}.  It was recently
shown~\cite{ak_mssm} that supersymmetric generalizations of the standard
model, in particular the MSSM, contain such solitons in their spectrum.
The role of the global symmetry in this case is taken by the U(1) symmetry
associated with the conservation of the baryon or lepton number.  Even more
remarkable is the fact that some of the Q-balls can be entirely stable
because their mass is less than that of a collection of nucleons with the
same baryon number~\cite{ks}.   

At the end of inflation in the early universe, the scalar fields develop
a large VEV along the flat directions of the scalar potential.  The
condensate  may carry some baryon or lepton number, in which case it can be
thought of as Q-matter, or a superhorizon-size Q-ball.  The subsequent
evolution of this condensate may give rise to the baryon asymmetry of the
universe~\cite{ad}.  However, a common assumption that the condensate
remains spatially homogeneous fails in many cases.  In fact, an initially
homogeneous solution of the equations of motion may become unstable with
respect to small coordinate-dependent perturbations~\cite{ks}.  The
exponentially fast growth of these perturbations can lead to fragmentation
of the scalar condensate with global charge into separate Q-balls.   Very
large stable baryonic Q-balls (B-balls) can be produced this way.

For most of my discussion I will not make any extra assumptions in addition
to those that lead to low-energy supersymmetry and inflationary cosmology. 
More specifically, the relations between the key conclusions and the
underlying assumptions can be illustrated by the following diagram: 

\vspace{3mm} 

\psfig{figure=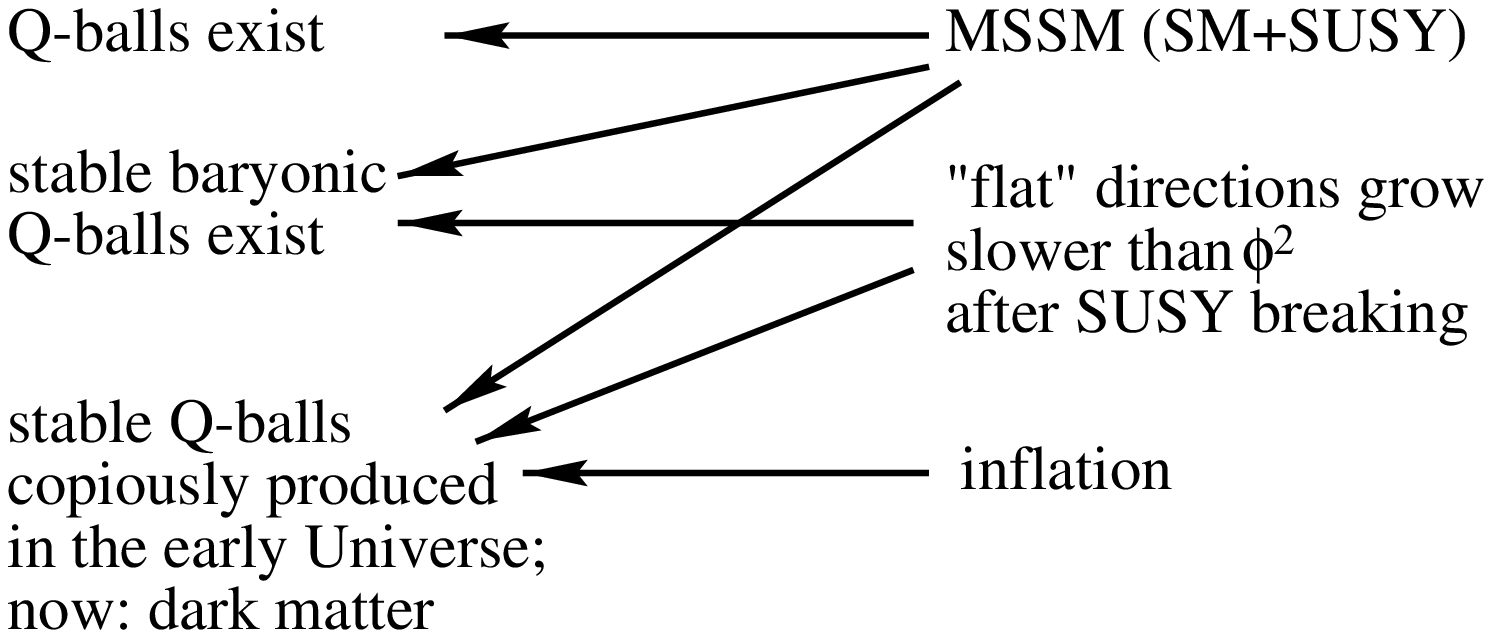,height=1.6in,width=4.0in}

\vspace{3mm} 

\section{Q-balls}

For a simple example, let us consider a field theory with a scalar
potential $U(\vf) $  that has a global minimum $U(0)=0$ at $\vf=0$.
Let $U(\vf)$ have an unbroken global\footnote{
Q-balls associated with a local symmetry have been constructed 
\cite{l}.  An important qualitative difference is that, in the case of a
local symmetry, there is an upper limit on the charge of a stable Q-ball.} 
U(1) symmetry at the origin, $\vf=0$.  And let the scalar field $\vf$ have
a unit charge with respect to this U(1).

The charge of some field configuration $\vf(x,t)$ is  
\beq
Q= \frac{1}{2i} \int \vf^* \stackrel{\leftrightarrow}{\partial}_t  
\vf \, d^3x . 
\label{Qt}
\eeq
Since a  trivial configuration $\vf(x)\equiv 0$ has zero charge, the
solution that minimizes the energy, 
\beq
E=\int d^3x \ \left [ \frac{1}{2} |\dot{\vf}|^2+
\frac{1}{2} |\nabla \vf|^2 
+U(\vf) \right], 
\label{e}
\eeq
and has a given charge $Q>0$, must differ from zero in some (finite)
domain.  This is a Q-ball.   It is a time-dependent solution, more
precisely it has a time-dependent phase. However, all physical quantities
are time-independent.  Of course, we have not proven that such a 
``lump'' is finite, or that it has a lesser energy than the collection of
free particles with the same charge; neither is true for a general
potential.  A finite-size  Q-ball is a minimum of energy and is
stable with respect to decay into free $\vf$-particles if 
\beq
U(\vf) \left/ \vf^2 \right. = {\rm min},
\ \ {\rm for} \ 
\vf=\vf_0>0 .
\label{condmin}
\eeq

One can show that the equations of motion for a Q-ball in 3+1 dimensions
are equivalent to those for the bounce associated with tunneling 
in 3 Euclidean dimensions in an effective potential $\hat{U}_\omega
(\vf)= U(\vf) - (1/2) \omega^2 \vf^2$, where $\omega$ is such that it
extremizes~\cite{ak_qb}
\beq
{\cal E}_\omega = S_3(\omega) +\omega Q. 
\label{Ew}
\eeq
Here $S_3(\omega)$ is the three-dimensional Euclidean action of the bounce
in the potential $\hat{U}_\omega (\vf)$ shown in Figure~1.  The Q-ball
solution has the form 
\beq
\vf(x,t) = e^{i\omega t} \bar{\vf}(x),
\eeq
where $\bar{\vf}(x)$ is the bounce. 

The analogy with tunneling clarifies the meaning of condition 
(\ref{condmin}), which simply requires that there exist a value of 
$\omega$, for which $\hat{U}_\omega (\vf)$ is negative for some value of 
$\vf=\vf_0 \neq 0$ separated from the false vacuum by a barrier. 
This condition ensures the  existence of a bounce.  (Clearly, the
bounce does not exist if $\hat{U}_\omega (\vf) \ge 0$ for all $\vf$ because
there is nowhere to tunnel.)  

\begin{figure}
%\postscript{figure1.eps}{0.3}
\setlength{\epsfxsize}{3in}
\setlength{\epsfysize}{2.4in}
\centerline{\epsfbox{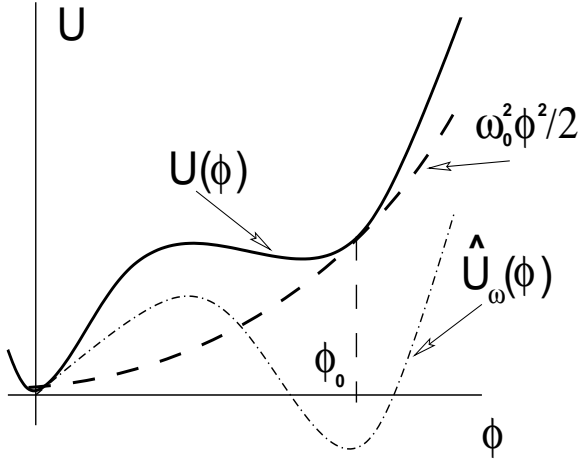}}
\caption{
The scalar potential {$U(\vf)$} (solid line) and the effective potential 
{$\hat{U}_\omega (\vf)$} (dash-dotted line) for some value of {$\omega$}. 
As charge increases, {$\omega$} decreases approaching {$\omega_0$}, the
coefficient of a parabola tangential to  {$U(\vf)$} (dashed line).    
} 
\label{fig1}
\end{figure}

In the true vacuum, there is a minimal value $\omega_0$, so that only for  
$\omega>\omega_0$, $\hat{U}_\omega (\vf)$ is somewhere negative (see
Figure~1).  If one considers a Q-ball in a metastable false vacuum, then  
$\omega_0=0$.  The mass of the $\vf$ particle is the upper bound on
$\omega$ in either case. Large values of $\omega$ correspond to small
charges~\cite{ak_qb}.  As $Q \rightarrow \infty$, $\omega \rightarrow 
\omega_0$.  In this case, the effective potential $\hat{U}_\omega (\vf)$
has two nearly-degenerate minima; and one can apply the thin-wall
approximation to calculate the Q-ball energy~\cite{q}.  For smaller 
charges, the thin-wall approximation breaks down, and one has to resort to
other methods~\cite{ak_qb}. 

The above discussion can be generalized to the case of several fields, 
$\vf_k$, with different charges, $q_k$~\cite{ak_mssm}.  Then the Q-ball is
a solution of the form 
\beq
\vf_k(x,t) = e^{iq_k \omega t} \vf_k(x),
\label{tsol}
\eeq
where $\vf(x)$ is again a three-dimensional bounce associated with
tunneling in the potential 
\beq
\hat{U}_\omega (\vf) = U(\vf)\ - \ \frac{1}{2} \omega^2 \, 
\sum_k q_k^2 \, |\vf_k|^2. 
\label{Uhat}
\eeq
As before, the value of $\omega$ is found by minimizing ${\cal E}_\omega$
in equation  (\ref{Ew}).  The bounce, and, therefore, the Q-ball, exists if 
\begin{eqnarray}
\mu^2 & = & 
2 U(\vf) \left/ \left (\sum_k q_k \vf_{k,0}^2 \right ) \right. = {\rm min},
\ \nonumber \\ & & {\rm for} \ |\vec{\vf}_0|^2 > 0.
\label{condmin1}
\end{eqnarray}

The soliton mass can be calculated by extremizing ${\cal E}_\omega$ in
equation (\ref{Ew}).  If $ |\vec{\vf}_0|^2 $ defined by equation
(\ref{condmin1}) is finite, then the mass of a soliton $M(Q)$ is
proportional to the first power of $Q$: 
\beq
M(Q) = \tilde{\mu} Q, \ \ {\rm if} \ |\vec{\vf}_0|^2 \neq \infty. 
\label{MQ}
\eeq
In particular, if $Q\rightarrow \infty$, $\tilde{\mu}\rightarrow \mu$
(thin-wall limit)~\cite{nts,q}.  For smaller values of $Q$, $\tilde{\mu}$
was computed in~\cite{ak_qb}.  In any case, $\tilde{\mu}$ is less than
the mass of the $\phi$ particle by definition (\ref{condmin1}). 

\begin{figure}
%\postscript{figure1.eps}{0.3}
\setlength{\epsfxsize}{3in}
\setlength{\epsfysize}{2.4in}
\centerline{\epsfbox{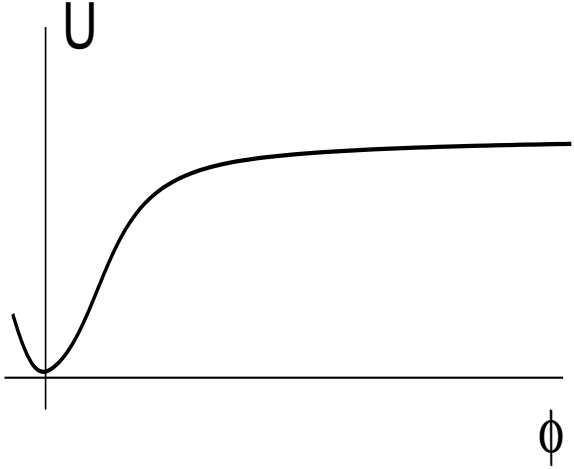}}
\caption{
An example of a ``flat potential'', for which {$U(\vf)/\vf^2$} is minimized
at $\vf=\infty$.  Such potential admits Q-balls, whose mass 
$M(Q) \sim m Q^{3/4}$ grows slower than the first power of $Q$. 
} 
\label{fig_flat}
\end{figure}

However, if the scalar potential grows slower than the second power of
$\phi$, then $|\vec{\vf}_0|^2 = \infty$, and the Q-ball never reaches the
thin-wall regime, even if Q is large.  The value of $\phi$ inside the
soliton extends as far as the gradient terms allow, and the mass of a
Q-ball is proportional to $Q^{p}$, $p<1$.  In particular, if the scalar
potential has a flat plateau $U(\phi) \sim m $ at large $\phi$, then the
mass of a Q-ball is~\cite{dks} 
\beq
M(Q) \sim m Q^{3/4}.
\label{MQflat}
\eeq
This is the case for the stable baryonic Q-balls in the MSSM discussed
below.

\section{Superballs in the MSSM}

The presence of the scalar fields with conserved global charges and the
requisite ``attractive'' interactions allows for the existence of Q-balls
in the supersymmetric extensions of the standard model.   Superpartners of
quarks and leptons carry the baryon and the lepton numbers that play the
role of charge $Q$ discussed above.  

There are two different sources of the attractive scalar self-interaction
that satisfy the criterion (\ref{condmin1}).  First, the tri-linear
couplings arise from the superpotential 
\beq
W=y H_2 \F \f + \tilde{\mu} H_1 H_2 +...
\label{sptn}
\eeq
as well as from supersymmetry breaking terms.  Here $\F$ stands for either
a left-handed quark ($\Q$), or a  
lepton ($\Lp$) superfield, and $\f$ denotes the right-handed $\q$ or $\lp$,
respectively.   The corresponding scalar potential must, therefore, have
cubic terms of the form  $y \tilde{\mu} H_2 \F \f$.  In addition, there are
soft supersymmetry breaking terms of the form $y A H_1 \F \f$.  The
condition (\ref{condmin1}) is automatically satisfied unless some Yukawa
couplings and some soft supersymmetry breaking terms are set to 
zero~\cite{ak_mssm}.  Therefore, Q-balls associated with baryon  
(B) and lepton (L) number conservation are generically present in the
MSSM.  The Q-balls associated with the trilinear couplings are
generally unstable and decay into fermions, quarks and leptons, in a way
similar to that discussed in Ref.~\cite{ccgm}. 

Another source of ``attraction'' that makes the condition  (\ref{condmin1})
possible is the flat directions in the MSSM.  Some gauge-singlet
combinations of the squarks and sleptons parameterize a number of
``valleys'' along which the scalar potential would have been zero were it
not for supersymmetry breaking. To avoid the problematic supertrace relation, 
it is commonly assumed that the supersymmetry breaking takes place 
in some hidden sector, that is, the sector that has no
direct couplings to the quark and lepton superfields in the superpotential.
The role of this sector is to provide a superfield $X$ (usually a singlets
under the standard model group) with a nonvanishing scalar and auxiliary
($F_X$) components. This breaks supersymmetry, and also ensures that no
unbroken $R$-symmetry survives. 
 The transmission of the supersymmetry breaking to the observable sector
is due to some messenger interaction with a typical scale $M_{_S}$. 
Supergravity, or some heavy particles charged under the
standard model gauge group, can be the messengers in the so called 
gravity-mediated or gauge-mediated scenarios, respectively. 
Integrating out the messenger sector below the scale $M_{_S}$, one is left
with the higher dimensional couplings (suppressed by powers of
$M_{_S}^{-1}$) between the observable and the hidden sector
superfields. The resulting scale of supersymmetry breaking in the
observable sector is set by the ratio $F_X/M_{_S}$. In this scenario the
soft masses are "hard" below the  scale $M_{_S}$ but they disappear above
that scale.  In the absence of detailed understanding of the origin of
supersymmetry breaking, I treat the scale $M_{_S}$ as a phenomenological
parameter that can be as low as several TeV (in gauge-mediated scenarios,
for example), or as high as the Planck scale.  In what follows we will
concentrate on the case in which $M_{_S}$ is below the scalar VEV in a
Q-ball.  This allows for stable baryonic Q-balls (B-balls) in the MSSM.  

In addition to global charges, the same scalars carry some gauge
charges as well.  The gauge structure of Q-balls is discussed in
Ref.~\cite{kst}.  If the effect of the
gauge fields cannot be eliminated, the semiclassical description of
the solitons may be hampered by the complications related to
confinement and other aspects of gauge dynamics.  In many cases, however,
one can construct a Q-ball using a gauge-invariant scalar condensate. 
This is true, in particular, in the MSSM, where all fields that have
non-zero VEV along the flat directions are necessarily gauge singlets. 

The mass of a Q-ball with a scalar VEV that extends beyond $M_{_S}$ along
some flat direction is determined by formula (\ref{MQflat}).  If the
condensate has a non-zero baryon number, the mass per unit baryon number
decreases with $Q_{_B}$, the baryon number of a B-ball: 
\beq
\frac{M(Q_{_B})}{Q_{_B}} = \frac{m}{Q_{_B}^{1/4}} < {\rm 1~GeV \ \ \ for \
  \ }  Q_{_B}> 10^{12} \left ( \frac{m}{{\rm 1~TeV}} \right )^4. 
\label{stable}
\eeq
A B-ball with a baryon number $Q_{_B}> 10^{12}$ is entirely stable
because it is lighter than a collection of neutrons and protons with
the same baryon number.  

If such large B-balls have formed in the early universe, they would
presently exist as a form of dark matter.  

In the early Universe, Q-balls can be created in the course of
a phase transition  (``solitogenesis'')~\cite{s_gen}, or they can be
produced via fusion~\cite{gk} in a process reminiscent of the big bang
nucleosynthesis (``solitosynthesis'').  However, it is unlikely that either
of these processes could lead to a formation of solitons with such an 
enormous charge.  

Very large Q-balls can form, however, from the breakdown of a primordial
scalar condensate~\cite{ks} that forms naturally at the end of inflation
and is  the key element of the Affleck--Dine baryogenesis.

\section{Fragmentation of the Affleck--Dine condensate} 

At the end of inflation, the scalar fields acquire large expectation values
along the flat directions.  Evolution of a scalar condensate carrying a
baryon or lepton number has been studied extensively in connection with the
Affleck-Dine scenario for baryogenesis in the MSSM~\cite{ad}.  However, a
commonly made assumption that an initially spatially-homogeneous condensate
remains homogeneous throughout its evolution turns out to be
wrong~\cite{ks}.  In fact, the condensate often develops an instability
with respect to small $x$-dependent perturbations that lead to
fragmentation of the condensate into Q-balls with the same types of global
charges.   

Indeed, the baryonic condensate of the form $\phi = e^{i \omega t} \phi_0$
is nothing but  Q-matter, or a universe filled with a Q-ball of infinite
size.  In a static universe, such field configurations are known to break up
into finite-size Q-balls under some conditions~\cite{klee}.  The expansion
of the universe makes the analyses more complicated. 

\begin{figure}
\setlength{\epsfxsize}{3.4in}
\centerline{\epsfbox{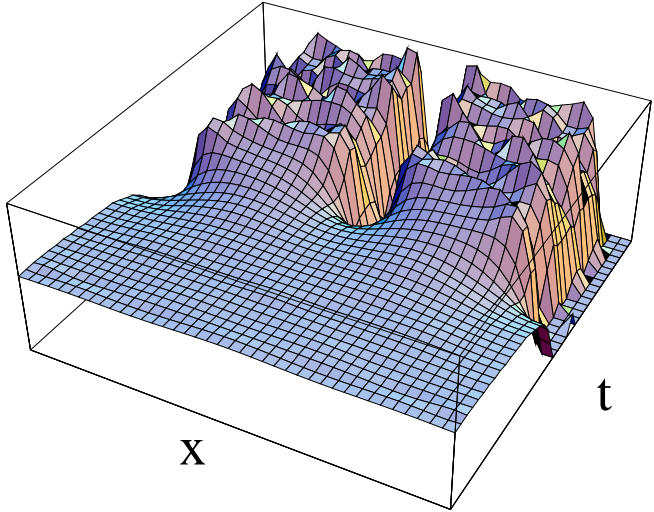}}
\caption{ 
The charge density per comoving volume in (1+1) dimensions for a sample
potential analyzed numerically during the fragmentation of the condensate
into Q-balls. 
}
\label{fig_charge}
\end{figure}

One can analyze the stability of a given slowly varying solution $\phi=R(t)
e^{i\Omega(t)}$ (where $R$ and $\Omega$ are both real) of the equations of
motion with a scalar potential $U(\phi)$ by adding a small space-dependent
perturbation $\delta R, \, \delta \Omega \propto e^{S(t) - i
  \vec{k}\vec{x}}$.  Then one can look for growing modes, ${\rm Re} \,
\alpha>0$, where $\alpha = dS/dt $.  The
value of $k$ is the spectral index in the comoving frame and is
red-shifted with respect to the physical wavenumber in the expanding
background: $\tilde{k}=k/a(t)$, where $a(t)$ is the scale factor. 

Of course, if the instability develops,  the linear approximation soon
ceases to be 
valid. However, we assume that the wavelength of the fastest-growing mode
sets the scale for the high and low density domains that collapse into
Q-ball.  This assumption can be verified {\it post factum} by comparison
with a numerical analysis, in which both large and small  
perturbations are taken into account.  

From the equations of motion one can
derive a dispersion relation that defines the band of unstable modes, 
$0<k<k_{max}$, where
\beq
k_{max}(t)=
a(t) \sqrt{\dot \Omega^2-U''( R )}.
\label{band}
\eeq
The  amplification of a given mode $k$ is characterized by the exponential
of $S(k)=\int \alpha(k,t) dt$, and depends on how long the mode remains in
the band of instability before (and if) it is red-shifted away from the
amplification region. 

It is natural to identify the best-amplified mode (that with maximal
$S(k)$) with the size of a Q-ball formed as a fragment of the initial
condensate.  

The detailed analyses of fragmentation for some potentials can be found in
Refs.~\cite{ks,em,e}.  The evolution of the primordial condensate can be
summarized as follows: 

\pagebreak

\vspace{3mm}
\psfig{figure=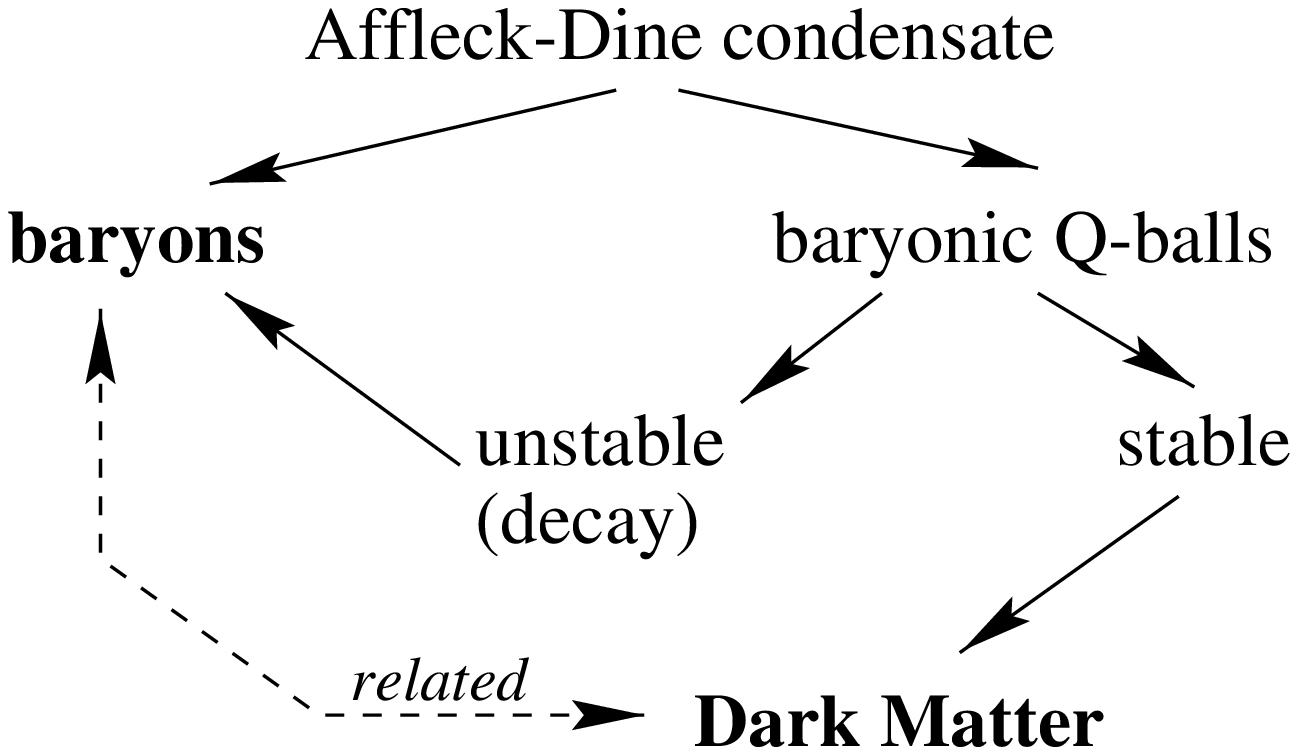,height=50mm,width=110mm}
\vspace{3mm}

Both the ordinary baryonic matter and the stable B-balls can be produced
from a single primordial scalar condensate.  Stable baryonic Q-balls make a
natural candidate for cold dark matter in theories with supersymmetry if
inflation took place in the early universe.  This scenario is particularly
appealing because, since the dark matter and the ordinary matter are
produced in the same process, their amounts are naturally related and are
calculable in a given model.  

\section{$\Omega_{nucleon}$ versus $\Omega_{_{DARK}}$}

Conceivably, the cold dark matter in the Universe can be made up entirely 
of superballs.  Since the baryonic matter and the dark matter share the
same origin in the scenario described in the previous section, their
contributions to the mass density of the Universe are related.  Therefore,
it is easy to understand why the observations find $\Omega_{_{DARK}} \sim
\Omega_{nucleon} $ within an order of magnitude.  This fact is extremely
difficult to explain in models that invoke a dark-matter  candidate whose
present-day abundance is determined by the process of freeze-out,
independent of baryogenesis.  If this is the case, one could expect
$\Omega_{_{DARK}}$ and $\Omega_{nucleon} $ to be different by many orders of
magnitude.  If one doesn't want to accept this equality as fortuitous, one
is forced to hypothesize some {\it ad hoc}  symmetries~\cite{kaplan} that
could relate the two quantities.   In the MSSM with AD baryogenesis, the
amounts of dark-matter Q-balls and the ordinary matter baryons are
naturally related~\cite{ks,lsh}.  One predicts~\cite{lsh} $\Omega_{_{DARK}} =
\Omega_{nucleon} $  for B-balls with  
\beq
Q_{_B} \sim 10^{26} \left ( \frac{m}{{\rm 1~TeV}} \right )^2.  
\label{Qdark}
\eeq

A different scenario that relates the amounts of baryonic and dark matter
in the Universe, and in which the dark-matter particles are produced from
the decay of unstable B-balls was proposed by Enqvist and
McDonald~\cite{em,em1}.  Kari Enqvist gave a review of this scenario in his
talk~\cite{e}.

\section{Detection of primordial superballs}

Interactions of the superballs with matter~\cite{dks,kkst} are determined by
the structure of the scalar condensate inside the Q-ball.   In the interior
of a B-ball the squarks have a large VEV and, therefore, the color SU(3)
symmetry is spontaneously broken (the Higgs phase).  The flat direction may
contain the sleptons and the Higgs fields in addition to the squarks. 
When a nucleon enters a Q-ball, it dissociates into quarks, and the 1~GeV
binding energy is released in soft pions.  Then quarks are absorbed into
the condensate.   Likewise, the electrons can be absorbed by a condensate 
in a $(B-L)$-ball, for example.  A Q-ball that absorbs protons and
electrons at roughly the same rate would catalyze numerous proton decays
on its passage though matter. 

However, the electrons cannot penetrate inside those
Q-balls, whose scalar VEV gives them a large mass.  For example, the 
simultaneously large VEV's of both the left-handed ($L_e$) and the
right-handed ($e$) selectrons along the flat direction give rise
to a large electron mass through mixing with the gauginos.  The  locked out
electrons can form bound states in the Coulomb field of the (now
electrically charged) soliton.  The resulting system is similar to an atom
with an enormously heavy nucleus.  Based on their ability to retain  
electric charge\footnote{It should be stressed that the condensate inside the
  Q-ball is electrically neutral (and it is also a singlet with respect to
  all non-abelian gauge groups)~\cite{kst}.   The electric charge is acquired
  through interactions with matter~\cite{kkst}.}, the relic solitons can be
separated in two classes: Supersymmetric Electrically Neutral Solitons
(SENS) and Supersymmetric Electrically Charged Solitons (SECS).  The
interactions of Q-balls with matter,  and, hence, the modes of their
detection, differ depending on whether the dark matter comprises SENS or
SECS.   

First, the Coulomb barrier can prevent the absorption of the 
incoming nuclei by SECS.  A Q-ball with baryon number $Q_{_B}$ and
electric charge $Z_{_Q}$ cannot imbibe protons moving with velocity $v \sim
10^{-3} c$  if $Q_{_B} \stackrel{<}{_{\scriptstyle \sim}} 10^{29} Z_{_Q}^4
(m/1\, {\rm TeV})^4$.  Second, the scattering cross-section of an
electrically charged Q-ball passing through matter is now determined by,
roughly, the Bohr's  radius, rather than the Q-ball size: $\sigma \sim \pi
r_{_B}^2 \sim 10^{-16} cm^2$. 

By numerical coincidence, the total energy released 
per unit length of the track in the medium of density $\rho$ is, roughly,
the same for SENS and SECS,
$dE/dl \sim $ $100 \, (\rho/1 \, {\rm g\,  cm}^{-3}) \,$ ${\rm GeV/cm}$. 
However, the former takes in nuclei and emits pions, while the latter 
dissipates its energy in collisions with the matter atoms.  

The overall features of the superball track are similar to those of the
rare Pamir event described in Ref.~\cite{pamir}.  (An optimist may consider
this a candidate event.) 

Assuming that superballs make up an order-one fraction of dark matter, one
can predict the number density 
\begin{equation}
n_{_Q}\sim \frac{\rho_{_{DM}}}{M_{_Q}} \sim 5 \times 10^{-5} \, Q_{_B}^{-3/4} 
\left ( \frac{1 {\rm TeV}}{m}
\right ) {\rm cm}^{-3}.
\label{num_dens}
\end{equation}
and the flux 
\beq
F\simeq (1/4\pi) n_{_Q} v \sim  10^{2} \, Q_{_B}^{-3/4} \left ( \frac{1 {\rm
      TeV}}{m} \right ) {\rm cm}^{-2} {\rm s}^{-1} {\rm sr}^{-1}
\eeq
of the dark-matter Q-balls~\cite{kkst}.  Given the predicted size of
dark-matter superballs (\ref{Qdark}), a passage of a Q-ball though any of
the presently operating detectors would be a very rare event.  For example,
for $Q_{_B}\sim 10^{26}$ and $m\sim {\rm 1~TeV}$, Super-Kamiokande would
see one event in a hundred years.  Of course, smaller Q-balls with baryon
numbers $10^{22}...10^{24}$ may be detected.   

Non-observation of superballs sets the limit on their baryon number
(assuming $\Omega_{_Q} \sim 1$).  The present limits on SECS comes from the
MACRO search for ``nuclearites''~\cite{dg}, which have
similar interactions with matter: 
$F < 1.1 \times 10^{-14}$ cm$^{-2}$~s$^{-1}$~sr$^{-1}$.  This translates
into the lower limit on the baryon number of dark-matter Q-balls, $Q_{_B} 
\stackrel{>}{_{\scriptstyle \sim}} 10^{21}$.  Signatures of SENS are 
similar to those expected from the Grand Unified monopoles that catalyze the
proton decay.  If one translates the current experimental limits from
Baikal~\cite{baikal} on the monopole flux, one can set a limit on the charge 
of SENS, $Q_{_B} \stackrel{>}{_{\scriptstyle \sim}} 3 \times 10^{22}$, for
$m=1~{\rm TeV}$.  Non-observation 
of Q-balls at the Super-Kamiokande after a year of running would improve
this limit by two orders of magnitude.   Of course, this does not
preclude the existence of smaller Q-balls with lower abundances that 
give negligible contribution to the matter density of the universe. 

Electrically charged Q-balls with a smaller baryon number can dissipate
energy so efficiently  that they may never reach the detector. 
SECS with baryon number $Q_{_B} \stackrel{<}{_{\scriptstyle \sim}}
10^{13} (m/1\, {\rm TeV})^{-4/3}$ can be stopped by the 1000 m of
water equivalent matter shielding.  Such solitons could not have been
observed by the underground detectors.  Therefore, in the window of
$Q_{_B}\sim  10^{12}...10^{13}$ the flux of SECS appears to be virtually
unconstrained.

The present limits will be improved by the future experiments, for example, 
AMANDA, ANTARES, and others.  A low-sensitivity but large-area (several
square kilometers) detector could cover the entire cosmologically
interesting range of $Q_{_B}$.  

\section{Star wreck: the superball invasion}

In non-supersymmetric theories, nuclear matter of neutron stars is the
lowest-energy state with a given baryon number\footnote{I remind the reader
  that black holes do not have a well-defined baryon number.}.  In 
supersymmetric theories, however, a Q-ball with baryon number $10^{57}$ can
be lighter than a neutron star.  I am going to describe a process that 
can transform a neutron star into a very large B-ball.  The time scale
involved is naturally of the order of billion years.

Dark-matter superballs pass through the ordinary stars and planets with a
negligible 
change in their velocity.  However, both SECS and SENS stop in the neutron
stars and accumulate there~\cite{sw}.  As soon as the first Q-ball is
captured by a neutron star, it sinks to the center and begins to absorb the
baryons into the condensate.  High baryon density inside a neutron star 
makes this absorption very efficient, and the B-ball grows at the rate that 
increases with time due to the gradual increase in the surface area.  
After some time, the additional dark-matter Q-balls that fall onto the
neutron star make only a negligible contribution to the growth of the
central Q-ball~\cite{sw}.  So, the fate of the neutron star is sealed when
it captures the first superball.   

According to the discussion in section 3, the energy per unit baryon number
inside the relic B-ball is less than that in nuclear matter.   Therefore,
the absorption process is accompanied by the emission of heat carried away  
by neutrinos and photons.   We estimate that this heating is too weak to
lead to any observable consequences.  However, the absorption of nuclear
matter by a baryonic Q-ball causes a gradual decrease in the mass of the
neutron star.  

Neutron stars are stable in some range of masses.  In particular, there is
a minimal mass (about 0.18 solar mass), below which the force of gravity is
not strong enough to prevent the neutrons from decaying into protons and
electrons.  While the star is being consumed by a superball, its mass
gradually decreases, reaching the critical value eventually.  At that
point, a mini-supernova explosion occurs~\cite{minisn}, which can be
observable.  Perhaps, the observed gamma-ray bursts may originate from an
event of this type.  A small geometrical size of a neutron star and a large
energy release may help reconcile the brightness of the gamma-ray bursts
with their short duration. 

Depending on the MSSM parameters, the lifetime of a neutron star $t_s$ can
range from 0.01 Gyr to more than 10 Gyr~\cite{sw}:
\beq
t_s \sim \frac{1}{\beta} \times 
\left ( \frac{m}{200\, {\rm GeV}} \right )^{5} {\rm Gyr}, 
\eeq
where $\beta$ is some model-dependent quantity expected to be of order
one~\cite{sw}.   The ages of pulsars set the limit $t_s> 0.1$~Gyr.   

It is interesting to note that $t_s$ depends on the fifth power of the
mass parameter $m$ associated with supersymmetry breaking.  If the 
mini-supernovae are observed (or if the connection with gamma-ray bursts is
firmly established), one can set strict constraints on the supersymmetry
breaking sector from the rate of neutron star explosions. 

The natural billion-year time scale is intriguing.

\section{Conclusion}
 
In conclusion, supersymmetric extensions of the standard model predicts the
existence of non-topological solitons, Q-balls.   They could be
produced in the early universe as a by-product of the Affleck -- Dine
baryogenesis.  Large baryonic superballs can be entirely stable and can
contribute to dark matter at present time.   This makes superballs a
natural candidate for dark matter in theories with low-energy
supersymmetry.  

Present experimental and astrophysical limits are consistent with superball
dark matter.  The relic Q-balls can be discovered by existing 
Baryonic Q-balls have strong interactions with matter and 
can be detected in present or future experiments.  
Observational signatures of the baryonic solitons are characterized by a
substantial energy release along a straight track with no attenuation
throughout the detector. The present experimental
lower bound on the baryon number $Q_{_B} \stackrel{>}{_{\scriptstyle \sim}}
10^{22}$ is consistent  with theoretical expectations~\cite{ks} for the
cosmologically interesting range of Q-balls in dark matter.   
In addition, smaller Q-balls, with the abundances much lower than that in 
equation (\ref{num_dens}), can be present in the universe.  Although
their contribution to $\Omega_{_{DM}}$ is negligible, 
their detection could help unveil the history 
of the universe in the early post-inflationary epoch.  Since the
fragmentation of a coherent scalar condensate~\cite{ks} is the only
conceivable mechanism that could lead to the formation of Q-balls with
large global charges,  the observation of any Q-balls would seem to speak
unambiguously  in favor of such process having taken place.  This would, in
turn, have far-reaching implications for understanding the origin of the
baryon asymmetry of the universe, for the theory of inflation, and for  
cosmology in general. 
 
The entire cosmologically interesting range of dark-matter superballs could
be covered by a detector with a surface area of several square kilometers.  
Since the required sensitivity is extremely low (thanks to the huge energy
release expected from the passage of a superball), it is conceivable that a
relatively inexpensive dedicated experiment could perform the exhaustive
search and ultimately discover or rule out superballs as dark-matter
particles.

\section*{Acknowledgements}

I am grateful to the organizers of DARK-98 for their hospitality.  Much of
the work I discussed was done in collaboration with M.~Shaposhnikov, as
well as G.~Dvali, V.~A.~Kuzmin, P.~G.~Tinyakov, and I.~I.~Tkachev. I thank
S.~Coleman and S.~Glashow for our recent discussions of SUSY Q-balls and
their detection.   Also, I thank J.~March-Russell who coined the name {\it
  superball}.

\end{document}